# TOWARDS AN ARTIST-CENTRED AI

**GORDAN KREKOVIĆ**
Visage Technologies
gordank@gmail.com

**ANTONIO POŠĆIĆ**
antonio.poscic@gmail.com

**DEJAN GRBA**
Interdisciplinary Graduate Center,
University of the Arts in Belgrade
dejan.grba@gmail.com

**ABSTRACT**

Awareness about the immense impact that artificial intelligence (AI) might have or already has made on the social, economic, political, and cultural realities of our world has become part of the mainstream public discourse. Attributes such as ethical, responsible, or explainable emerge as associative and descriptive nominal references in guidelines that influence perspectives on AI application and development. This paper contextualizes the notions of suitability and desirability of principles, practices, and tools related to the use of AI in the arts. The result is a framework drafted as a set of atomic attributes that summarize the values of AI deemed important for artistic creativity. It was composed by examining the challenges that AI poses to art production, distribution, consumption, and monetization. Considering the differentiating potentials of AI and taking a perspective aside from the purely technical ontology, we argue that artistically pertinent AI should be unexpected, diversified, affordant, and evolvable.

# 1. INTRODUCTION

Artificial intelligence (AI) is one of the most rapidly expanding branches of computer science. It has been applied in a swath of scientific disciplines and industries, studied by academia, portrayed in popular culture, and covered by the media. Similar to other novel technologies, it foments oppositions between techno-optimists and Luddites and stirs disparate opinions and schools of thought about its conceptual nature, technical issues, and development. Its contradictions and tensions, ranging from different types of mystifications to corporate or government-driven abuses, are widely addressed by recent research.[1] During the past decade, critics have become vocal about systemic dangers tied to the concepts of artificial autonomy and entrepreneurial deployment of AI systems, pointing out David Collingridge's caution (1980) that in this phase the technology develops much faster than society can comprehend and regulate (Johnson and Verdicchio, 2017; Tegmark, 2017). "Ethical", "responsible", "explainable", and "decolonial" are some of the ethically desirable features that have been increasingly explored in AI discourse with regard to the applied AI's extensive reach and socioeconomic influence (Jobin et al., 2019). On the opposite side of the critical spectrum, some researchers and pundits see AI as a route toward automated social organization and transhumanist utopia (Kurzweil, 2005; Fridman, 2022).

In the context of the arts, AI has come to prominence through two vectors. One includes tech-savvy media artists who made creative use of the accessibility of machine learning (ML) architectures and tools (Miller, 2019; Cetinić and She, 2022, pp. 10-11). The other comprises commercial products and free sandbox-style tools that allow the generation of similes via user-friendly interfaces (Steinbrück, 2021; Roose, 2022). Analogously to the reactions to AI in other fields, this development has been seen either as a milestone in the history of art or as a harbinger of its death. With some exceptions (Audry, 2021), media art theory and philosophy of art studies address modern AI phenomenology primarily by describing the characteristics of existing technologies and treating them as black boxes with specific inputs and outputs (Manovich, 2019; Żylińska, 2020). In this paper, we take the perspective of the actual and potential implications of AI on artistic creativity by stepping out of technological ontology. Inspired by the discourse on AI ethics, characteristics, and modes of application, which recommends policies and guidelines for desirable AI, we outline key principles that could constitute a framework for artistically desirable AI. The framework comprises four atomic, yet descriptive and associative attributes that summarize the values of AI that may be important for artistic practice.[2]

We start with a quick overview of modern AI's ethical considerations. Focusing on the factors of expressive authenticity, we proceed toward the uses of AI in artistic practices and look at their cultural implications, philosophical inferences, and the aspects in which technological research aims to augment or enrich artistic creation. In the central section, we discuss AI as a tool that, beyond being widely affordable, can be (critically) deconstructed and remodelled according to the artists' needs and requirements. Based on these observations, we ask: if there are universal guidelines for building and using AI systems for artistic creation, what would be their widely desirable characteristics? To trace a path toward viable answers, we identify a set of AI's related attributes and try to ascertain how to change and implement them in order to make the technology suitable for a broad range of artistic purposes.

[1] The corpus of publications includes Cathy O'Neil's accounts of the injustices and fallouts of unregulated application of big-data-based AI (2016), Max Tegmark's examination of the challenges of future AI (2017); the conceptual, technological, and sociocultural critique of AI science, technology, and business by Melanie Mitchell (2019) and Gary Marcus and Ernest Davis (2019); Michael Kearns and Aaron Roth's (2019) discussions of the sociopolitical and cultural questions opened by the AI's technical imperfections and biases; studies on the AI-affected future of work, edited by Benedict Dellot et al. (2019); the assessment of AI's political and ideological realms by Nick Dyer-Witheford et al. (2019); Brian Christian's (2020) survey of the value/interest alignment problems in machine learning; Michael Betancourt's (2020) study of the cultural impacts of AI; and Kate Crawford's (2021) mapping of the exploitative layers of AI research and business..

[2] Related efforts include AI Art Manifesto by Natalia Fuchs et al. (2020), AI Art Field Guide by the Partnership on AI (2020), and *Creating AI Art Responsibly: A Field Guide for Artists* by Claire Leibowicz et al. (2021).

## 2. DESIRABLE AI

Ethical questions about the coexistence, interaction, and mutual influence between artificial agents and their creators arose in the early depictions of autonomous systems in science fiction literature (Hermann, 2021). The word *robot* was first introduced in Karel Čapek's play R.U.R. (Rossum's Universal Robots, published in 1920) which thematizes the issue of non-humans becoming a servant class of human society. In *I, Robot*, published around twenty years later, Isaac Asimov introduced the Three Laws of Robotics as a set of rules that define basic principles of robotic behaviour or, in Asimov's words (1981), "the only way in which rational human beings can deal with robots". Under the related premise that artificial agency is subjected to the design choices of its creators so that understanding and anticipating the consequences of these choices is essential for responsibility assessment, ethics has been an integral part of AI research since its outset in the mid-1950s.

In a meta-analysis of 84 guidelines for ethical AI, Jobin et al. (2019) identified eleven clusters of principles: transparency, justice and fairness, non-maleficence, responsibility, privacy, beneficence, freedom and autonomy, trust, sustainability, dignity, and solidarity. As a necessity-driven contemporary elaboration of Asimov's Three Laws, this comprehensive set of attributes illustrates how experts and ethicists see an AI that can fit broader civilizational values. Similar to other regulatory guidelines and policies, these recommendations are broad, diverse, demanding, and important but remain inconsistently adopted and implemented. Notional discrepancies in ethical values merge with ideologies and economic power interests to form a complex sociopolitical environment that accommodates the majority of affairs and debates about AI.

As an industrial technology, AI has been increasingly applied to automate tasks and jobs traditionally reserved for human cognitive capabilities, with potentially vast socioeconomic, political, and cultural consequences. AI is entangled with aggravating factors such as algorithmic biases introduced by training machine learning (ML) models on historically biased data (Veale and Binns, 2017), the lack of interpretability (Burrell, 2016), and the liability and intellectual property issues (Zeilinger, 2021). These problems are exacerbated by the interconnectedness and interdependency of the global economy in which the reach and impacts of applied AI systems can be massive. Various ways in which "neutral mathematical tools" can affect individual destinies and collective life may be disturbing when clearly described (O'Neil, 2016). The applied AI's consequentiality is further complicated by the ambiguous functionality of ML systems: in some contexts, they are useful and controllable, and in others superfluous, absurd, or abusive. For instance, recommendation algorithms can improve the user experience of digital products but when they optimize for profitability or displace other means of content access, they become detrimental to perception and decision-making. Tomo Kihara's online work *TheirTube* (2020) (Kihara, 2020) warns about these implications. It is a filter bubble simulator that generates YouTube landing pages with recommended videos for six fictional profiles (Fruitarian, Prepper, Liberal, Conservative, Conspiracist, and Climate Denier), based on their previous YouTube viewing data.

Concerns that AI may be misused by malevolent actors (Brundage 2018), jeopardize human jobs (Nature, 2017; Dellot et al., 2019), or undermine social fairness through systemically biased or erroneous inference (Zou et al., 2018) have been raised in scientific and mainstream discourse. Consequently, governments, non-profit organizations, professional associations, as well as the private sector, have been constituting expert committees tasked to draft AI policies and guidelines. These include the High-Level Expert Group on Artificial Intelligence appointed by the European Commission, the expert group on AI in Society of the Organisation for Economic Co-operation and Development (OECD), the select committee on Artificial Intelligence of the United Kingdom House of Lords, the Association of Computing Machinery (ACM), Amnesty International, and companies such as SAP and Google.

Motivated by these broad issues, we aim to understand and describe the desirable features of AI in the context of artmaking. Within a range of the applied AI's influences on the artworld, our focus is on the impacts on artistic creativity, expressivity, and poetic identity and on the cultural positioning of art. To sketch the desirable features of AI in such contexts, it is useful to identify its potential risks and evidently negative, detrimental, limiting, or otherwise undesirable effects or implications.

## 3. AI AS ART'S EXISTENTIAL CRISIS

Although it does not directly involve most professional artists, the AI's "black-box" effect could be highly consequential for them in the long term. Regardless of artistic skillsets, art-historical knowledge, or understanding of the underlying technologies, casual users of systems such as ChatGPT, Make-a-Scene, DALL·E 2, GLIDE, Imagen, Parti, Simulacrabot, Dreamstudio, or Midjourney can make textual prompts or simple parameter modifications to output ostensibly original artefacts that may be identified as artworks (Marcus et al. 2022). This machinic emulation of certain types of artistic processes sheds interesting light on an old question of the ratio between an artist's comprehension of their means and the poetic cogency of their work. With limited insight into the functionality of their tools, how clearly can artists formulate expressive intentions, make creative choices, and define their role in ideation, construction, presentation, communication, and other complex social and relational processes of artmaking? Whether commercial or freely available, many ML systems for augmented creativity are designed around particular usage schemas that impose specific procedural and aesthetic vocabularies (Carnovalini and Rodà, 2020). With the reduced space for the artist's direct influence, both the aesthetic and epistemological directions in the development of an AI artwork are significantly determined by the opaque heuristics of their tools, which do not necessarily correspond with the poetically relevant values (Manovich, 2018). This may lead to stylistic homogenization and turn creative production into a reinforcement of technologically embedded (political) interests and biases comparable with the filter bubbles blown by the taste-inferring algorithms that govern streaming services (Bello and Garcia, 2021; Chen and He, 2021).

At the same time, the AI-powered mimicry of recognizable artforms and styles creates a phantom of social agency and creative autonomy that devalues human-made works and induces false but widespread fear that artists will be supplanted by algorithms (Browne, 2022) or that automation will radicalize the precarity in most areas of professional artmaking (Kral et al., 2019). As a combined effect, some artists may become apprehensive of AI and perceive it as a threat to their practices, livelihoods, and raison d'être (Moruzzi, 2020). They may reject the technology in principle, regardless of its potential for exploring new avenues of expression. The fact is that the dominant intents and purposes of AI development are traditionally driven by scientific, technological, and economic interests so designing deeply competent and highly customized artistic tools is not their main focus. Aside from being used for the promotional purposes of the corporate sector, artmaking resides on the margins of AI techno-solutionism, mainly because of its relatively low economic exploitability. Therefore, many original artworks are created by hacking, customizing, repurposing, diverting, or subverting AI tools that were not built specifically for making art. That is one of the reasons that many artists still largely focus on exploring the technical aspects of the concepts and real-world manifestations of ML (Grba, 2022, pp. 19-20) resulting in similar ideas and narratives that establish a broadly recognizable expressive identity, which may become misleadingly synonymous with AI art as a whole.

A related set of problems stems from AI's impact on larger systems for art production, distribution, consumption, and monetization. Subscription-based streaming services such as Spotify and Netflix redefined the creative mechanics and aesthetics of music, film, and television by transforming the methods of their dissemination and reception. These companies continuously build and reshape intricate business models around suites of AI algorithms that maximize engagement and link user interaction to various commercial offers (Finn, 2017). By favouring forms, contents, and styles most lucrative within such operational logic, they inevitably manipulate collective taste and define the ecosystems of both consuming and making music and film/TV.[3] While their creative demands drive some artists away (Chai, 2022), their financial power and market dominance attract the majority of others and gradually nurture a new consumerist monoculture (Gaw, 2022). Intentional or unintentional cultural colonialism in ML design, training, and application introduces another reinforcement factor of expressive homogenization. Most ML training datasets are assembled from Internet-scraped data which is saturated by few globally dominant cultures so the resulting models will implicitly favour these cultures' languages and tropes and become better versed in replicating their underlying values on account of non-hegemonic cultures (Koutsomichalis and Achilleos, 2021).

These issues partly coincide and overlap with the crises of (digital) artmaking initiated by the technical logic and trends in the crypto art market (Zeilinger, 2018). Crypto art regime shifts the focus of appraisal from the artistic process as an intrinsically non-capitalist type of work (Beech, 2015) onto the transaction-centred entrepreneurship, which renders creative production vulnerable to exploitation and regresses the creative ethos of digital art into the obsolete notional orbits of possession and unique "ownership". In direct contradiction to the copyability of the digital medium, the NFTs facilitate a dubious imposition of faux rareness or a fictitious "crypto-aura" onto artefacts whose "aura" should emanate from their poetic, expressive, and relational values. They institute a ghost of property or an imaginary experience of ownership, which should not be confused with "aura" because there is no such thing as an authentic digital copy (Juárez, 2021; McCormack, 2022). By restricting the key poetic features of digital art, such as accessibility, mutability, and interchangeability, this commodification logic undermines a range of experimental and critical practices that have been emancipating the notion of an artwork from a sacred (or fetishized) material entity toward a relational process of ideational exchange.

## 4. ETHICS OF AI IN ART

The use of AI technologies under the umbrella notion of creative expression by a wide variety of practitioners—from amateurs to engineers and scientists to professional artists in different genres and niches—raises a range of ethical concerns about the biases, inequities, and injustices in AI technologies, the environmental impact of AI development and application, and the AI-powered economic, decisional, and executive power aggregation in the hands of the owners of large computational infrastructures.

Beyond that techno-social substrate, AI art makers (in all categories) are ethically liable when their efforts yield vacuous works, indulge in derivative aesthetics that seduces or entrances viewers into cultural conformity and political deference, overlook AI's socio-political context, dismiss artistic for technical competencies or downplay important technological aspects, ignore their works' poetic similarities with other, more convincing works, or avoid disclosing and unacknowledging them (Grba, 2023a, pp. 211–213).

The appreciation of modern AI's ethical issues is most evident in experimental and mainstream AI art practices where some actors believe that the only ethical artistic use of AI technologies is to address their socio-political, economic, and ethical issues. However, even well-intended and competently produced tactical efforts are susceptible to ethically charged slippages. Works that aim to contest AI biases toward certain ethnic, gender, and social groups sometimes end up exploiting their cultural contexts or communities they purport to protect, such as Jake Elwes' multipart *Zizi Project* (since 2019) (Grba 2022, p. 13). Even projects that explicitly critique AI's ethical flaws can themselves be ethically flawed, as exemplified by the technical errors and moral conflicts in Kate Crawford and Trevor Paglen's project *Training Humans* (2019–2020), which nevertheless gained the art world's wide acclaim for its cogent treatment of AI bias (Leibowicz et al. 2021, p. 7).

By finding loopholes and underscoring weaknesses in the AI systems they criticize, artists set their achievements up for recuperation, which allows the corporate sector to refine the normalization of injustices instead of correcting its political directives and improving its ethical standards. Even when intervening proactively, critical AI art opens questions beyond its apparent contributions. Should an activist action end up (directly or indirectly) being used by the AI industry to enhance its profitable instruments or remedy its public image without necessarily improving its techno-ethical standards? Could tactical art disrupt the corporate AI regime with lasting and desirable social consequences, and how effectively can it incite or enhance government policies for accountability and regulation of private AI businesses with global influence? Namely, both culturally and technically constrained critical attention to AI's ethical issues obscures their fundamental and often elusive sociopolitical facets, while the false allure of "solving" them detracts us from bigger, more pressing, and longer-term challenges (Powles, 2018). All these ethical issues combine with AI artists' intrinsic need for endorsement by the tech sector, academia, and art market.

However, rather than moralizing, the critique of AI art ethics should be constructive towards the betterment and maturation of the field. AI art's ethical problems reflect not only contemporary culture but also AI-related science, technology, economy, politics, and social relations and imply ethical responsibility in all these domains (Grba 2023b, pp. 511–513). Any critical artistic engagement with AI should problematize the political influence and social impact of the AI research communities, corporate sector, and art market and call for an actionable scepticism toward their ideologies and interests (Payne, 2013). Yet, while resources such as the *AI Artists' Manifesto* (Fuchs et al., 2020) or *Responsible AI Art Field Guide* (Leibowicz et al., 2021) provide useful ethical considerations and guidance, artists remain primarily responsible and should be held accountable for the sociocultural values of their production. In a range of contributions stretched between providing veneer and cultural legitimization to the corporate AI sector and taking genuine risks for cutting-edge expression, AI artists' potential to establish poetic cogency depends on their ability to cultivate well-informed ethical attitudes toward their professional goals and overcome the systemic but hazardous entanglements with corporate tech and the art market.

## 5. QUESTIONS OF CREATIVITY

The value of creativity in artmaking is as contextually dependent as in other areas, so technologies such as AI, which bring the artefact-producing tools to a wide community whose members don't necessarily have the technical skills or knowledge required for making these artefacts in other (older) media or technologies, does not democratize artmaking insomuch as it evolves the modes (practices) of artmaking, the notions of art, and the criteria of art appraisal. In a wider perspective, however, the availability of user-friendly AI-powered tools for making complex artefacts may be beneficial for stimulating their users' overall creativity and cognitive processes, although these longer-term gains would be difficult to assess objectively because, inter alia, the "creativity imperative" is one of the key exploitative incentives of the cognitive capitalism (Reckwitz, 2017).

## 6. ARTISTICALLY DESIRABLE AI

The discourse about best practices for desirable AI frames adjectives such as transparency, fairness, inclusiveness, responsibility, sustainability, trustfulness, etc., which are clarified and elaborated upon in the accompanying documentation or in the ensuing discussion. Following this approach, we draft a small set of artistically desirable values of AI taking into account basic requirements, challenges, and implications of contemporary artmaking. Although they reflect certain principles, these values are descriptive rather than prescriptive and open to adaptation, expansion, modification, or redaction according to individual interpretation and with regard to the inherently emergent character of the arts and AI technologies and the contingencies of their actualization.

The rapid development of AI technologies in general and the nascent nature of their use in arts imply potential uncertainties of the applicability of the desirable values outlined here for future evolutions. However, one of the factors that was considered in the value selection process was their intrinsic epistemological stability and pertinence impervious to changes in the underlying techno-social systems and avenues of cultural production. Furthermore, while the methodologies and processes of AI related tools are currently progressing at a fast pace – disregarding signs of stagnation (Mosqueira-Rey et al., 2023) and issues with explainability (Minh et al., 2023) – the overarching artistic and conceptual reference points in the field have remained stable (Boden, 1998) and rooted in qualities external to the technologies employed. The findings of this paper are therefore likely to remain applicable barring major paradigm shifts, which are inherently unpredictable.

## 6.1. UNEXPECTEDNESS

Since the launch of Google's *DeepDream* in 2015, the repertoire of corporate AI marketing strategies includes selecting some functional scopes of novel ML models, adorning them with user-friendly interfaces, and making them available online for public use. They usually become popular tools for transforming text, images, music, videos, and other media and contribute to the culturalization of AI through a proliferation of artefacts on social media and other platforms (Ferrari and McKelvey, 2022). Unlike traditional art technologies, systems such as Dall-E 2, Stable Diffusion, ChatGPT, or MuseNet do not require extensive training, experience, and time to achieve formally plausible results but their extensive use easily exhausts in uniformity despite the apparent variability of potential outputs. An early entry in the contemporary AI art spring— DeepDream—serves as a telling example.

In 2014, Google developed a convolutional neural network (CNN) for computer vision codenamed GoogLeNet and next year released its "Inception" module[4] under the title *DeepDream* as a software package that digital artists started exploring, and as an online app that became widely popular and turned into a digital meme generator (Tyka, 2015). *DeepDream* produces delirious visuals that merge pareidolia with mise en abyme and Droste effects in a quasi-style called Inceptionism (Szegedy, 2015). Inceptionist works by Mehmet Memo Akten, Josh Nimoy, Samim Winiger, Mike Tyka, and other artists and engineers caught the artworld's attention and were featured in several exhibitions, such as *DeepDream: The Art of Neural Networks* at San Francisco's Gray Area Foundation for the Arts in 2016. However, the fractal uniformity of psychedelic and hallucinatory effects tends to homogenize into a bland signature (Nimoy, 2015) so Inceptionism struggled to transcend formal experimentation, demonstration, or decorative intervention and the style quickly dried out. Besides the structural uniformity and formal similarities between Inceptionist works, the main reason is that arbitrary generation of mimetic imagery or animations becomes oversaturated and boring if it unfolds unbounded. In order to engage the viewer, it requires prudently defined conceptual, narrative, and formal constraints, which seemed to be difficult to implement with DeepDream. A tendency towards formal repetitiveness and homogenized aesthetics is evident across other areas of AI art such as GANism (Mira, 2019; Żylińska, 2020), ML-derivative mainstream practices, and large-scale projects (Grba, 2022, pp. 19-20).

In that context, the differentiating potential for AI in the arts lies in discovering ways to generate unexpected but meaningful forms, narratives, or events by leveraging both the unique technical characteristics of AI architectures and the specific contexts of their application. Possible approaches to attaining authentic expressive modes include conceptually strong prompting and curation of outputs, building custom training datasets, and subverting the use or deployment of existing architectures. Exemplars include Jake Elwes' *Closed Loop* (2017), François Quévillon's *Algorithmic Drive* (2018–2019), Bill Posters and Daniel Howe's *Big Dada: Public Faces* (2019–2021), Golan Levin and Lingdong Huang's *Ambigrammatic Figures* (2020), Terence Broad's *Teratome* (2020), Martin Disley's *How They Met Themselves* (2021), and various works by Jennifer Gradecki and Derek Curry, Nao Tokui, Sebastian Schmieg, Anna Ridler, Mario Klingemann, and Shinseungback Kimyonghun artist duo.

The common denominator of this attribute is in employing AI competently to extend the possibilities and transcend the expressive boundaries of conventional art techniques and traditional tools. It still requires significant technical expertise and time investment for skill building, experimentation, and data processing, but the expanding palette of solutions makes it gradually more approachable for creative exploration and learning. In general, the range of approaches to achieving expressive unexpectedness should not be exclusively dependent on the technical characteristics of any given architecture or on its novelty, so the caveat of this perspective is the potential fetishization of technically induced originality.[5]

## 6.2. DIVERSIFICATION

The aesthetic influence of creative tools does not emanate only from their technical capabilities, limitations, and impositions but also from the artists' inclinations and cultural contexts. An example is post-Internet art (sometimes also called post-digital art and post-media art), which centres around treating information technologies as common utilities for appropriating, referencing, and emphasizing the digital affects of culture and everyday life (Paul, 2015). Thanks to the increasing accessibility of digital creative tools, the post-Internet approach was embraced by professional artists but also spearheaded a number of subcultures and microgenres. The recognizable motifs of Inceptionism, GANism, deepfaking, and other expressive trends indicate similar patterns in AI art as a wider audience associates AI with the slight uncanniness of deepfaked imagery or with the formal signatures of GAN visuals: morphed patches with blurred areas, averaged colour textures, and regional variance of details and sharpness (Audry, 2021, pp. 163-166).

The AI's artistic potential lies in the ability to traverse and merge arbitrary expressive "vocabularies" and go beyond stylistic directives by combining AI with other (non-AI) tools and infrastructures. For instance, with *Muse AI Supercuts* (2017, commissioned for the rock band Muse), digital agency Branger_Briz extended the conceptual and technical logic of automated supercut[6] by intersecting ML with Web search (Grba, 2020, p. 68). They designed an ML system that assembled daily music videos in which every lyric of Muse's song *Dig Down* (2017) is voiced by a different notable person from YouTube videos. Another example is Allison Parrish's book of poems *Articulations* (2018, part of the series *Using Electricity*), which transposes the popular AI art method of latent space sampling (Cetinić and She, 2022, pp. 12-13) to exploit non-visual data through 5 Many AI artists are tech-savvy creators who value virtuosity in solving technical problems or making technical improvements over other fundamental aspects and, as Audry (2021, p. 184) admits: "it is common for artists to spend most of their time developing the technological infrastructure necessary to support their work, with only a fraction spent on artistic and aesthetic work—which often occurs in the last few days before a show". Looking hard for the "magic" into the depths of artificial neural layers, artists tend to be mesmerized by the sensations of ineffability and wonder of their self-organizing expression and risk missing the higher order of these intrinsic qualities within the totality of their creative processes. Instead of building novelty through unique ways of articulating neural networks in broader ideational, topical, narrative, or procedural frameworks, they content with prying mildly surprising outputs out of the unintelligible processes of deep neural networks. Consequently, their observations of expressive limits often collapse into the intrinsic constraints (features and bugs) of the techniques they use.

6 Supercut is an edited set of short media sequences selected and extracted from their sources by at least one recognizable criterion. By focusing on the specific elements (words, phrases, scene blockings, visual compositions, shot dynamics, etc.), supercuts can accentuate the formal repetitiveness and clichés in film, television, music, literature, common speech, etc.

smart manipulations of the phonetic features of words from a training dataset.

This type of versatility allows artists to avoid stylistic crystallization or to escape merging with mainstream or pop-cultural memetics. While striving for unexpectedness may lead to a convergence into unique associative motifs that gradually become predictable or clichéd, artists also have the freedom to continuously unthink, undefine, and reinvent their expressive language through the intersective diversification of conceptual, methodological, formal, and aesthetic spaces.

## 6.3. AFFORDANCE

The growing popularity of online marketplaces for ML models, such as SingularityNET, GenesisAI, AWS Marketplace, Nuance AI Marketplace, or Bonseyes, has prompted the research and development communities to start shaping up best practices that integrate tools, automation, infrastructures, and operational workflows (MLOps) to help programmers shift their focus from building models to applying them. The development of no-code, drag-and-drop ML platforms such as Microsoft's Power Apps environment, Automated ML and AutoML on Google Cloud, Amazon Web Services mobile and web app builders, Apple's Create ML and MakeML, targets business clients who want to use ML with minimal skills or insight into the code (Webb et al., 2022). This platform service trend, although selective in protecting the owner/providers' corporate interests, will probably induce some level of trickle-down effect with consumer-grade, user-friendly AI applications in other areas, including art. Pre-trained and trainable ML models, knowledge bases, and skill-building resources are available to artists in various forms: as data science notebooks, such as Google Collab or Jupyter Notebooks, stored in codebase repositories such as Open Model Zoo, GitHub, and Hugging Face, as specialized frameworks and libraries, as web tools such as Magenta's, in standalone software plug-ins, or in mobile apps.

The variance in technical accessibility of these resources configures a landscape whose cognitive requirements and creative scopes may be of equal interest to the artists. As is the case with other artistic tools, the expressivity of AI instruments relies on their affordances and on the breadth of outcomes they may produce in the hands of tech-savvy players, beginners, experimentalists, or casual users. Thus, the efforts on demystifying the technological aspects of ML tools, making them accessible and understandable to various categories of creators, and exposing their controllability on multiple levels are crucial for diverse AI artmaking. Nevertheless, despite AI's explosive and often chaotic development, it unfolds in parallel with the processes of standardization and crystallization of instruments, services, usage scenarios, and other options, which may limit the range of AI's affordances for artistic purposes.

## 6.4. EVOLVABILITY

The pace of enhancing the existing and developing new AI techniques and the scope and efficiency of their implementation should be leveraged in artistic practices and acknowledged in the art systems. As new versatile architectures, more efficient models, data acquisition/processing systems, and user-friendly tools become accessible, the artists' ideational, procedural, and presentational scopes need to expand proportionally and broaden the expressive range. Such technological and cultural dynamics require the symmetrical adaptability of unexpectedness, diversification, and affordance to new realities.

However, the speed of technological change may get in opposition to various types of other complexities and requirements of artmaking. In that regard, deepfaking may serve as a cautionary case of how not only the techniques but also the concepts, themes, and forms of AI art are being digested into pop-cultural products at raising speed. For instance, Libby Heaney's interactive karaoke installation *Resurrection (TOTB)* (2019), which intersects face-swapping deepfakes with pop music, was followed after between a few months and two years by apps such as Wombo (lip-syncing), Reface (face swapping), MyHeritage Deep Nostalgia™ (animation of photographs), or Jiggy (dancing). In fact, the open-source package Faceswap was first committed to GitHub on 15 December 2017 (Faceswap 2021), thus preceding *Resurrection (TOTB)* and several other deepfake artworks based on face swapping.[7]

On the other hand, the agility of the artists' inclusion of new AI techniques in their creative repertoires may collide with the institutional inertia of art representation, consumption, and education. For instance, Disnovation.org's project *Predictive Art Bot* (since 2017) uses natural language processing to question the discursive authorities and curatorial paradigms of contemporary art. It is a chatbot that generates concepts for art projects based on current art discourse and occasionally prophesizes absurd future trajectories for art on its own website and on Twitter. Another witty exemplification of core value discrepancies is Guido Segni's *Demand Full Laziness: A Five-Year Plan* (2018-2023). Segni delegated part of his five-year artistic production to a set of GANs that record his periods of inactivity or "unproductivity", such as sleeping, reading, or lazing about, and generate new images which are distributed to his Patreon sponsors. This ironic take on the trend of automating and "optimizing" artistic output integrates the contemporary reality of precarious labour with the critique of popular notions about the influence of AI on the identity of artmaking, the routinized aesthetics of GANism, and the (AI) artists' opportunism.

Depending on the developmental vectors of AI technologies and artmaking, different domains of AI art will in some circumstances resonate with the institutional contexts and oppose them in others. Largely acting on the level of individuals or small groups, the artists' evolvability should be expected to outpace the institutions, which may motivate better organizational fluency and more just operational flexibility of the art system or help institutions and educational systems to adapt more efficiently, 7 The closing gap of novelty does not necessarily undermine the epistemological value of well-conceived artworks but the ever-increasing temporal proximity to consumerism compromises their poetic authority and cultural identity. For a comparison, while the stylistic signatures of modernist painters such as Juan Miró or Piet Mondrian had easily found their way into the decorative design patterns of post-Second World War consumer culture, the self-cancelling aesthetics of Marcel Duchamp's readymades remains to this day too radical to be assimilated in that manner, even though it has been influencing several generations of artists and reverberating in their work (Molderings, 2010).

but it may also foster widening of the existing dichotomies such as those between mainstream and experimental art or between the academia and do-it-yourself culture.

## 7. CONCLUSION

The artistically desirable features of AI and the surrounding concerns outlined in this framework draft (as summarized in Figure 1) are not exhaustive but serve as initial reference points for further consideration, discussion, study, and refinement. An important factor for its elaboration is the awareness of the fluid and always potentially deceptive conceptual nature of its central terms: art, creativity, and intelligence, as well as the sociotechnical dynamics of AI. Intrinsically, art, creativity, and intelligence have no clear or consensual definitions and constantly redefine themselves. AI research, technological development, and applications change quickly, often with sweeping consequences. They all contribute to the real-world complex systems of economy, technology, and society, which are inherently unpredictable (Tetlock and Gardner, 2015). Therefore, the parameters of this framework should be empowered by synergizing the sense of temporality with vigilance and caution to anticipate and articulate the artistic potentials, trade-offs, issues, or dangers of the emergent AI and AI-related techniques and instruments.

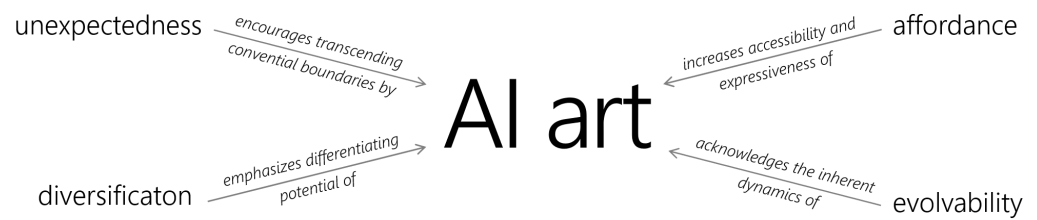

Figure 1. Artistically desirable values of AI. © Authors

The four attributes we discussed may implicate an overarching dichotomy between engineering practices in developing ML systems and artistic approaches to working with these systems. Modern sub-symbolic ML design revolves around the "problem-solving" approach and optimization enhancement logic. Operating in an incommensurable space of possibilities, artists often seek to pose questions and expose problems rather than solve them, usually avoid optimization, and do not always respond directly to the audience's inclinations, tastes, and preferences (Audry, 2021, pp. 56-59). While it is important for artists and scientists to acknowledge and understand these differences, they are not crucial in principle because most technologies were not initially designed or intended for artmaking, but artists adopt, adapt, combine, and transform them for expressive purposes. That is exactly the case with successful AI artists who use ML both as a rich unifying complex of productive methods and as a contextual domain. Their lessons should be integral to the calibration of this framework through self-reflection and reckoning not only about what is desired but who desires what and why.